*sensors*

*Article*

# Custom IMU-Based Wearable System for Robust 2.4 GHz Wireless Human Body Parts Orientation Tracking and 3D Movement Visualization on an Avatar

Javier González-Alonso [1,*], David Oviedo-Pastor [1], Héctor J. Aguado [2], Francisco J. Díaz-Pernas [1], David González-Ortega [1] and Mario Martínez-Zarzuela [1,*]

1    Grupo de Telemática e Imagen, Universidad de Valladolid, 47011 Valladolid, Spain;
     moraleja39@gmail.com (D.O.-P.); pacper@tel.uva.es (F.J.D.-P.); davgon@tel.uva.es (D.G.-O.)
2    Unidad de Traumatología, Hospital Clínico Universitario de Valladolid, 47003 Valladolid, Spain;
     hjaguado@gmail.com
*    Correspondence: javier.galonso@alumnos.uva.es (J.G.-A.); marmar@tel.uva.es (M.M.-Z.)

**Abstract:** Recent studies confirm the applicability of Inertial Measurement Unit (IMU)-based systems for human motion analysis. Notwithstanding, high-end IMU-based commercial solutions are yet too expensive and complex to democratize their use among a wide range of potential users. Less featured entry-level commercial solutions are being introduced in the market, trying to fill this gap, but still present some limitations that need to be overcome. At the same time, there is a growing number of scientific papers using not commercial, but custom do-it-yourself IMU-based systems in medical and sports applications. Even though these solutions can help to popularize the use of this technology, they have more limited features and the description on how to design and build them from scratch is yet too scarce in the literature. The aim of this work is two-fold: (1) Proving the feasibility of building an affordable custom solution aimed at simultaneous multiple body parts orientation tracking; while providing a detailed bottom-up description of the required hardware, tools, and mathematical operations to estimate and represent 3D movement in real-time. (2) Showing how the introduction of a custom 2.4 GHz communication protocol including a channel hopping strategy can address some of the current communication limitations of entry-level commercial solutions. The proposed system can be used for wireless real-time human body parts orientation tracking with up to 10 custom sensors, at least at 50 Hz. In addition, it provides a more reliable motion data acquisition in Bluetooth and Wi-Fi crowded environments, where the use of entry-level commercial solutions might be unfeasible. This system can be used as a groundwork for developing affordable human motion analysis solutions that do not require an accurate kinematic analysis.

**Keywords:** body tracking; inertial sensor; wearable sensor; custom; do-it-yourself; motion capture

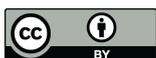

## 1. Introduction

In recent years, the development and validation of motion capture systems based on the use of IMUs (Inertial Measurement Units) have been the subject of growing interest [1–3]. There has been an evolution from the use of single-isolated sensors [4], such as accelerometers and gyroscopes, towards the use of more complex single-pack solutions including an additional magnetometer and sensor fusion algorithms that provide more accurate movement data [1] and reduce sensor drift [5–7].

Broadly speaking, the research on wearable inertial and magnetic sensors for human motion analysis includes works focusing on quantifying human movements and works focusing on movement classification [1]. The requirements and the outcome of the different systems vary depending on the target application. Some studies concentrate their efforts on obtaining the most accurate kinematic analysis of the movement [8], which is useful for identifying biomechanical disorders or atypical patterns of neuromuscular control [9].

IMU-based systems have proven to be useful for assessing upper limb movements [10–13], lower limb movements [14–16], and gait [3,15,17,18]. The applicability of these solutions is supported by comparative precision studies with optoelectronic systems, considered the gold standard in movement acquisition [19], traditional measurement tools such as goniometers [20], isokinetic dynamometers [14,21], or custom designed measurement tools using servomotors for static and dynamic testing [19]. Other studies are focused on well-balanced systems that can provide a good enough estimation of the movement and higher-level interpretation. These systems find their applicability in rehabilitation [22–26] and motion assessment [27–30], even in the natural environment of the patient.

Despite the benefits and potential uses of this technology, there are some barriers that need to be overcome before they can be widely employed in real life applications. The more featured and expensive high-commercial solutions, such as Xsens MTw Awinda [31], Perception Neuron [32], Opal (APDM) [16,19,33], or Physiolog (Gait Up) [34] are not being routinely used in the clinical settings and are far from being used in the natural environment of the patient. This is either motivated by the high price of the hardware, the need to pay additional user licenses for movement analysis tools, or their complexity of use.

Significantly, less featured entry-level commercial versions of IMU-based systems, such as Xsens DOT and Notch, are now available [35,36]. Unfortunately, they still present some limitations. First, the software provides only raw data, without interpretation. These systems do not include any software for human movement visualization during acquisition. Consequently, they are not ready-to-use solutions for the interested newcomer. Their use requires a strong technical and mathematical knowledge for sensor calibration and raw data processing and analysis.

In the second place, they are limited in the number of sensors that can be connected: only five to six sensors can be used simultaneously, which reduces the number of target applications. Finally, they suffer reliability data acquisition issues in environments where several Wi-Fi and Bluetooth networks coexist [37,38].

It is worth noticing that, while entry-level solution applicability is to be improved, in academia there is a growing number of research papers studying the feasibility of using non-commercial custom-developed do-it-yourself IMU-based sensors in medical applications [14,17,18,39,40]. Even though custom-designed solutions can lead to a wide use of this technology, the description of how to build them from scratch is yet too scarce in the literature. In addition, although these proposals are promising, they are still very limited in terms of connectivity and usability. One of the main drawbacks of the systems presented in previously published works is the limited number of sensors that can be used simultaneously. Significantly, few works record the orientation of upper and lower limbs at the same time [1,41]. Furthermore, in some studies the wearables are wire connected [10,12] or do not provide real-time data streaming to another device, but store it in Secure Digital (SD) or internal memories [11].

The limitations of entry-level commercial systems and state-of-the-art custom IMU-based systems led us to develop a do-it-yourself wearable system for robust 2.4 GHz wireless multiple human body parts tracking. The aim of this paper is two-fold. Firstly, it proves the feasibility of building an affordable and easy to use solution for multiple bone orientation tracking; while providing a detailed bottom-up description of the required hardware, tools, and mathematical operations to estimate and represent 3D movement using a 3D avatar. Secondly, it shows how the introduction of a custom 2.4 GHz communication protocol, including a channel hopping strategy, can also address some of the current communication limitations of entry-level commercial solutions.

The detailed contributions of this work are: (1) An affordable custom-designed IMU-based wireless system for multiple body parts orientation tracking, comprising up to 10–12 wearable sensors; (2) A description of a simple IMU-to-Segment (I2S) approach using quaternion data and an easy-to-use interface for real-time data acquisition and 3D movement visualization on the Unity3D game engine; (3) A demonstration of the benefits of a channel hopping wireless communication strategy in the 2.4 GHz for a more robust communication between sensors, even in Bluetooth and Wi-Fi crowded environments.

The remainder of this paper is organized as follows. Section 2 describes the materials and methods used to build the proposed system. Section 3 describes the experiments conducted and the obtained results. Section 4 discusses the achieved results and compares our work with previous studies. Finally, Section 5 presents the main conclusions obtained.

## 2. Materials and Methods

### 2.1. Hardware

Our system consists of Custom Wearable (CW) devices containing an IMU, a data processing and transmission core, and a power Lithium Polymer (LiPo) 90 mAh battery (or, alternatively, a rechargeable button battery), all in the same package. The dimensions of a single CW sensor are 45 × 26 × 10 mm, similar to the Xsens DOT (XDOT) sensors (36 × 30 × 11 mm) as shown in Figure 1.

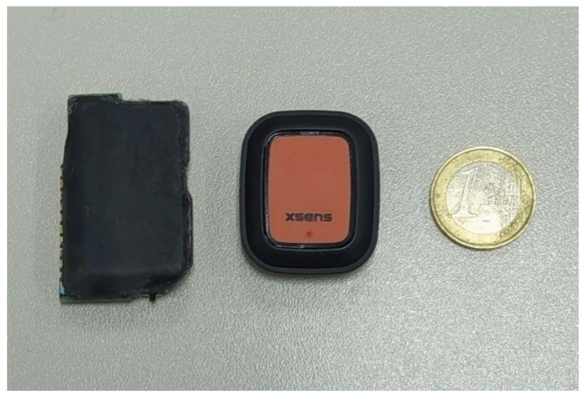

**Figure 1.** Relative size of the CW device and the XDOT sensors.

For prototyping the CW sensors, we used ARM Cortex M4 processors with a 2.4 GHz transceiver (approximate communication range of 7–8 m) from Nordic's nRF52 series, and BNO080 inertial sensors which provide good response to environmental disturbances [12]. This processor polls data from the IMU through a TWI peripheral interface (similar to I2C).

Table 1 compares the features of the BNO080 and the IMUs included in some high-end and entry-level commercial solutions. All of them include a triaxial accelerometer, a gyroscope, and magnetometer and, in general, the characteristics of the BNO080 are similar to its counterparts. In some cases, all the characteristics were not fully clarified by the manufacturer. The Nominal Dynamic Accuracy specifications of the BNO080 are the following: 0.3 m/s$^2$ for the accelerometer, 3.1°/s for the Gyroscope and 1.4 μT for the Magnetometer. Different IMU-based systems intended for kinematics analysis in real time have been described in the literature, either custom or commercial, concluding that the selection of specific sensor characteristics depends mainly on the final application and the joints involved [2,36]. For example, there are some works that focused on clinical gait analysis [42,43] or real-time monitoring of the rehabilitation process [44], which involve commercial systems including even ±6 g accelerometers (Opal version 2, from APDM). Among custom IMU-based solutions, also ±8 g accelerometers, such as the LSM303DLH, have been chosen for ROM assessment [45]. In sports, the specific activity may or may not require a higher grade solution, as proposed in "SoccerMate" [46] using an ActiGraph wGT3X-BT accelerometer with a range up to ±8 g, such as the BNO080. For this reason, the internal characteristics of the chosen sensor were considered optimal for our purposes.

Table 1. Features of internal IMU sensors of the proposed system and other commercial systems.

| Wearable Solution | Do-It-Yourself | Entry-Level | | High-End | |
|---|---|---|---|---|---|
| | Our CW | Xsens DOT | Notch | Xsens MTwAwinda | Perception Neuron PRO |
| Gyroscope | ±2000 dps | ±2000 dps | ±4000 dps | ±2000 dps | ±2000 dps |
| Accelerometer | ±8 g | ±16 g | ±32 g | ±16 g | ±16 g |
| Magnetometer | ±13 Gauss | ±8 Gauss | ±16 Gauss | ±1.9 Gauss | - |
| Dynamic Orientation error | 2.5–3.5° | 1–2° | 1–2° | 0.75–1.5° | 1–2° |

An important difference between the BNO080 and the rest of the IMUs is that it is packaged with an embedded ARM Cortex M0+, which provides in-chip data fusion. This BNO080 System in Package (SiP) also allows automatic calibration to occur in normal device use with no explicit user input. The BNO080 sensor reports include a specific field that informs the host microcontroller about the internal sensor's accuracy status [41]. Due to this internal procedure, the application knows the calibration status of the BNO080 and will request our attention on a further calibration if necessary. In this case, if a user wants to force a calibration, a series of common steps will have to be followed to calibrate the internal sensors. For the gyroscope, the device should be set down on a stationary surface for approximately 2–3 s. For the magnetometer, the device should be rotated about 180° and back to the beginning position for about 2 s in each axis (roll, pitch, and yaw). Regarding the calibration of the accelerometer, it needs to be done at least once after manufacture, by moving the device into four to six unique orientations for 1 s. Nevertheless, the final developed system is prepared to require no more calibration, and if necessary, this can be done easily by moving the storage box with the sensors inside.

This SiP provides different operation modes that can be set to modify the way the sensor estimates the orientation and uses the data provided by the accelerometer, the gyroscope, and the magnetometer. Rotation Vector is the most precise data fusion mode. In this mode, the BNO can reduce the drift by using the magnetometer and achieves a maximum static rotation error of 2.0° and a dynamic rotation error of 3.5° [47]. In terms of dynamic orientation error, commercial solutions present a slightly higher accuracy. Regardless, the BNO080 error is under 5°, which is considered the maximum accuracy error, as proposed by the American Medical Association for movement analysis in clinical situations [48].

### 2.2. Communication

The first layer of software in the proposed system is the firmware of the CW sensors. The firmware is programmed in C++, runs on the FreeRTOS operating system, and provides BNO080 data acquisition and wireless data communication. The data sent by a CW sensor is transmitted to a USB dongle connected to a computer in a master-slave scheme, as depicted in Figure 2. The master device polls all connected slaves at a maximum rate of 60 Hz for real-time movement data acquisition.

On the contrary, in the Xsens DOT communication schema, all sensor nodes establish a connection and send data to a device that must be BLE-capable. Xsens DOT are primarily designed to connect to mobile devices such as smartphones or tablets. In addition, the release of the official Xsens DOT server application enables the Xsens DOT communication with a computer, as shown in Figure 3.

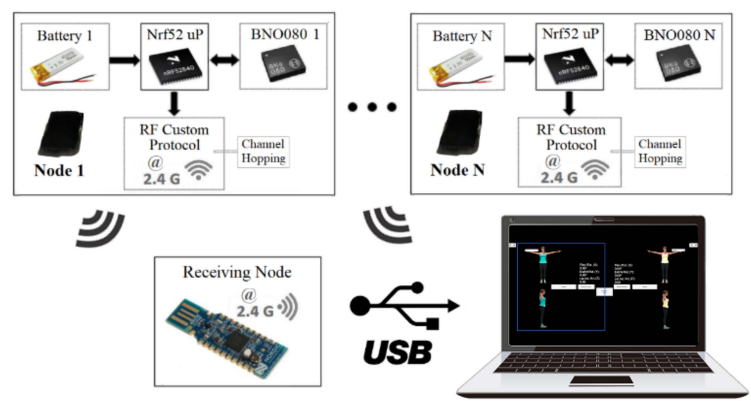

**Figure 2.** Communication between the CW sensors and the USB dongle attached to a computer.

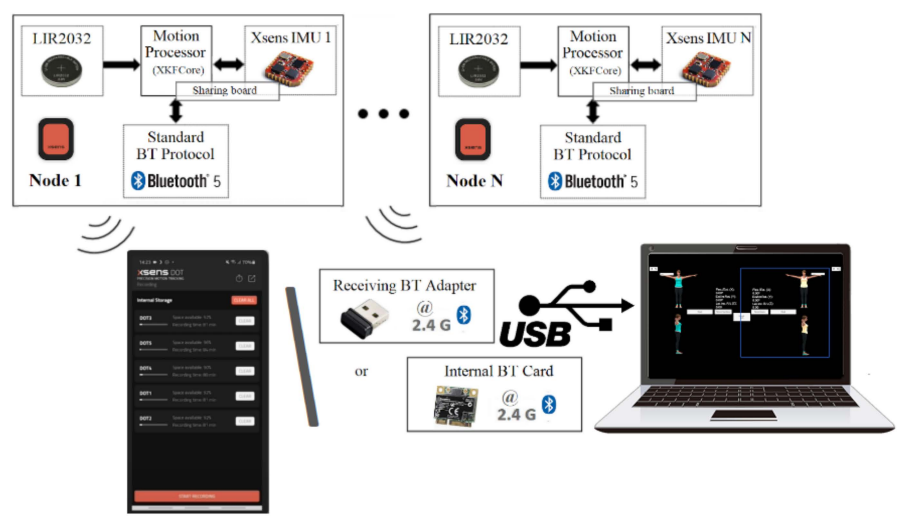

**Figure 3.** Communication between the Xsens DOT sensors and a BT-capable computer.

In addition to a different communication topology, in the proposed system a custom communication protocol was implemented. This protocol is implemented on top of the nRF52 on-chip low-energy wireless implementation for embedded devices [49], intended to operate in the Industrial-Scientific-Medical (ISM) 2.4 GHz frequency band. In the Physical Layer (PHY), Gaussian Frequency Shift Keying (GFSK) modulation is used just as in Bluetooth Low Energy (BLE) entry-level commercial solutions.

Although this BLE standard defines strategies to prevent interferences with other devices, its communication robustness can be affected in practice by other devices working in the 2.4 GHz band. Common industrial IoT specifications, such as Zigbee or other 802.15.4 implementations, may cause interferences in the sensors signal [50]. Data synchronization errors and interferences can make the communication unfeasible in these environments with a highly occupied band [31].

The proposed system includes a custom algorithm for improving communication robustness on the 2.4 GHz, increasing the final number of simultaneous sensors, reducing interference with other signals, and preventing data loss. This approach uses carrier frequencies different from those employed by the Wi-Fi and Bluetooth channels and implements a channel switching mechanism for the data transmission on top of our communication protocol. Specifically, it defines three fixed channels for connecting and synchronizing the CWs and 77 channels for data transmission. This way, the master and the slaves can switch to a different radio channel when an excessive number of lost packets or collisions is detected on a certain channel. A bandwidth of 2 MHz and an air data rate of 2 Mbps were used to provide lower average current consumption and reduce on-air collisions probability.

Coexistence between Bluetooth and Wireless LAN networks can be improved using a combination of Adaptive Frequency Hopping (AFH) and proprietary techniques for switching between the defined 40 standard BLE channels. However, the Xsens DOT BLE specifications do not clarify whether some techniques have been implemented to deal with this problem even using BT 5.0 capacities. Therefore, the feasibility of this entry-level solution should be verified based on experimental results conducted in environments with adverse conditions.

### 2.3. Body Parts Orientation Tracking and 3D Movement Visualization

The proposed system can communicate at a frequency in the range of 50–60 Hz with up to 10 CW sensors, allowing human movement data acquisition of the main joints of the body shown in Figure 4a. For real-time sensor data acquisition and 3D movement visualization, an application was developed using a free version of the Unity3D game engine. This second layer of software has been designed to represent the movements of the user wearing the CW or even the XDOT sensors on a standard rigged Unity3D avatar, as shown in Figure 4b.

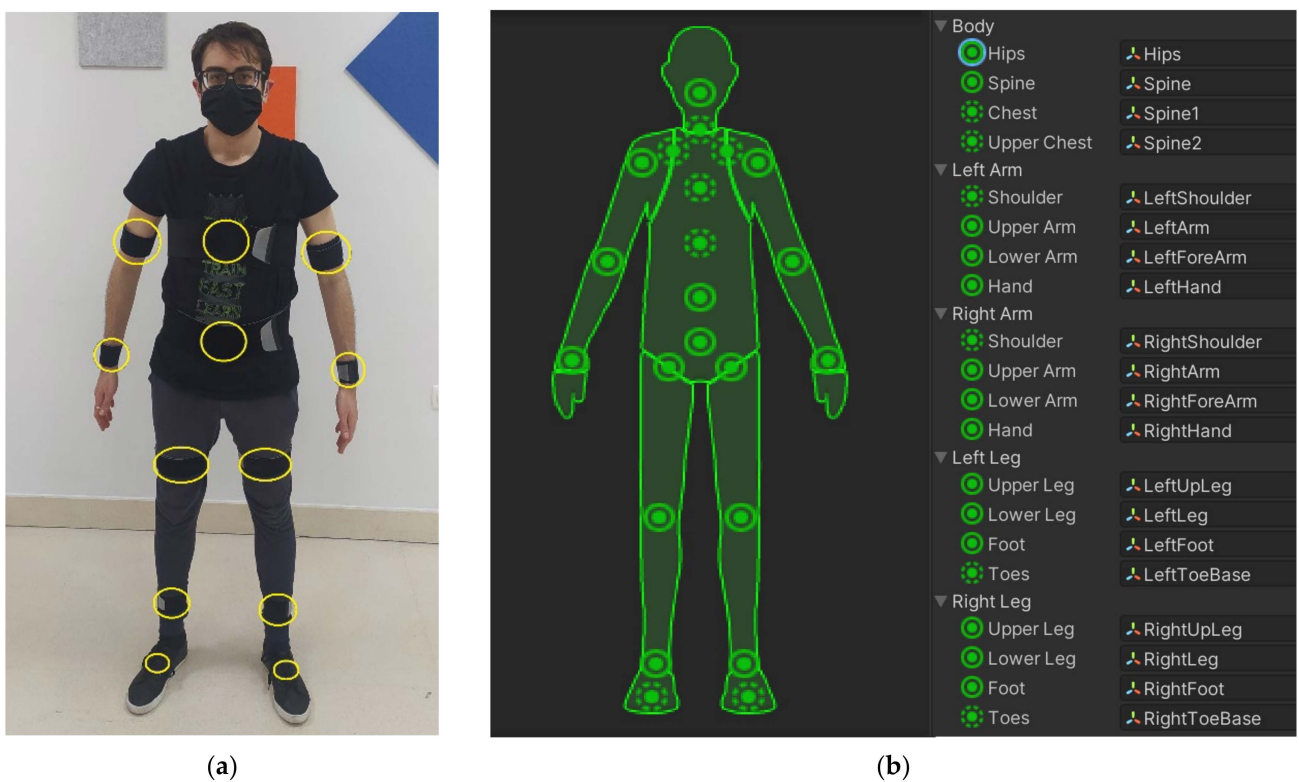

(**a**)           (**b**)

**Figure 4.** (**a**) Custom Wearable (CW) sensors placement on the human body. 12 sensors configuration: (1) back, (2) arm, (2) forearm, (1) pelvis/pubis, (2) thigh, (2) calf, and (2) foot. Back and pelvis sensors are placed rear-faced and in the backside of the straps (hidden in the image) and feet sensors are placed on the uppers, over, or underneath the shoelaces. (**b**) Avatar bone mapping, including the 12 bones in (**a**), plus: (1) chest, (1) upper chest, (2) shoulder, (2) hand/wrist, (2) toes.

The output of Custom Wearable (CW) devices is in ENU local earth-fixed reference coordinate system, a common standard in inertial navigation for aviation and geodesic applications. This system is defined as X positive to the East (E), Y positive to the North (N) and Z positive when pointing up (U), as depicted in Figure 5. When the CW sensor nodes are powered on, a heading reset is automatically performed. This must be done while they are allocated in their storage box, so that they are placed in parallel with each other.

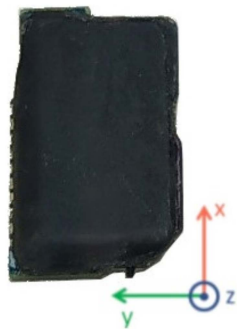

**Figure 5.** Custom Wearables right-handed coordinate system.

The CW sensors register their instantaneous 3D orientation and acceleration. The orientation is stored as a quaternion (Equation (1)), a complex number with three imaginary components that can encode an axis-angle rotation about an arbitrary axis in space. In the proposed system, we take a similar approach.

$$q = w + xi + yj + zk = (q_0, q_1, q_2, q_3) \tag{1}$$

In IMU-based movement tracking systems, the human body is represented as a biomechanical model consisting of segments that are assumed to be rigid and connected by ideal joints with a varying degree of freedom (DoF) [51]. Hinge joints such as the knee have one major rotation axis and limited range of motion (1 DoF). Other joints, such as the elbow, ankle, and wrist can be modeled using two non-perpendicular revolute joints (2 DoFs). More complex joints can be modelled as ball-and-socket joints (3 DoFs), where a capsule represents the soft tissue that links connected segments [52]. The proposed solution in this paper uses a standard rigged Unity3D avatar that is built-up using Unity3D configurable joints, which can be set up to force and restrict rigid body movement in any DoF. The bones considered in the 3D model are 20, as depicted in Figure 4b. Using this standard model eases the procedure of CW sensor to bone mapping. The bones to be animated depend on the number of available sensors for data capturing.

In the literature, we find two clearly different approaches to the estimation and analysis of human movement: *straightforward estimation* and *combined estimation* [2]. In the *straightforward estimation*, data offered by the accelerometer, gyro, and magnetometer is combined through a sensor fusion algorithm, measuring the orientation with respect to a global reference system, and then attaching each sensor to a specific body segment, acquiring a common reference system. Thus, the absolute orientation of a body segment in space is estimated independently from the adjacent segments. The *combined estimation* approach incorporates one or more of the anatomical constraints at the same time embedded in the estimation model, including surface geometry and soft-tissue constraints. This approach provides basic information of human motion from joint kinematics models by analyzing the position or orientation of a particular joint.

In this work, we parametrize each segment in the global space by its orientation [53], being represented by a unit quaternion. Using quaternion data, we introduce a simple yet very effective IMU-to-segment (I2S) calibration method useful for 3D movement visualization. During I2S, each sensor is assigned to its corresponding virtual body segment, considering both the rotation of the sensor around the real body segment and the initial pose of the subject, which should match the initial pose of the 3D avatar in the screen as much as possible. In the presented system, we propose taking advantage of quaternion properties to simplify the I2S operation, as follows:

1. The subject is asked to adopt a calibration body pose (the Unity3D app offers the user to adopt a neutral pose or a T-pose as Figure 6 shows). The instant rotation of the attached CW sensor is stored as the initial orientation $q_{calib}$. Then, the new relative orientation of the sensor in the IMUs global space $q'$, with respect to its initial orientation, can be computed using equation (2), where $q_{calib}^{-1}$ denotes the inverse of the initial orientation of the sensor in the human body pose adopted during the I2S procedure.

$$q' = q * q_{calib}^{-1} \qquad (2)$$

2. Since the CW sensors and the Unity3D environment do not use the same coordinate reference system, a translation is needed to adapt the raw data acquired by the sensors to the left-handed system adopted in the Unity3D application, using Equation (3), where $q''$ is now the relative orientation of the IMU in the Unity3D space.

$$q'' = (q'_0, q'_2, -q'_3, -q'_1) \qquad (3)$$

3. During the calibration procedure, the initial rotation of the humanoid bones $q_{bone}$ in the chosen initial pose is also considered. The avatar can be animated in real-time by rotating the humanoid bones according to Equation (3), in which $r$ denotes the instant rotation of the bone in the Unity3D global coordinate system.

$$r = q'' * q_{bone} \qquad (4)$$

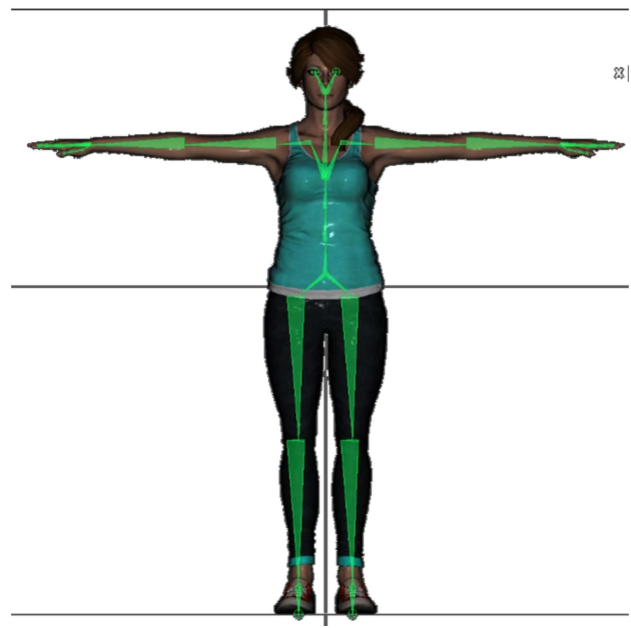

**Figure 6.** Avatar model in Unity3D and underlying skeleton in T-pose.

As previously commented in Section 2.2., the Unity3D application can also interface with the official Xsens DOT Server, what enables compatibility with these entry-level IMU sensors. Motion data is stored in a comma-separated values (CSV) file, which includes timestamps and quaternion data coming from every sensor. Data files can be used afterwards to reproduce the movement offline on a 3D avatar or for further analysis using Jupyter notebooks. These notebooks are programmed in Python, make use of the *scipy* module for quaternion operations, and *matplotlib* for graph plotting. They can use as input both our CW system data files and Xsens DOT data files, whether they are CSV files from our custom Unity3D program or those generated by the Xsens DOT official Android App.

### 2.4. Limitations of the Chosen Kinematic Model

The kinematic model employed in the proposed system does consider the orientations of the CW sensors with respect to the bones, but not their relative positions. Although we consider this approach to be valid for some applications, it has the limitation that only bone rotations can be estimated and the final position of body parts in a 3D space is unknown. The pelvis or the backbone can be chosen to be the reference joint. This reference joint can rotate in 3D, but its absolute position in the 3D space cannot be modified (e.g., a user walking will appear walking in a static location in the app). On the other hand, the proposed solution provides a very fast calibration method, and it is yet possible to estimate joint angles as the relative orientation between bones.

In applications aiming at greater accuracy in movement analysis, a complete spatial description of the multibody kinematic model should be developed. That full kinematic model would be inherently application dependent [54] and requires the definition of personalized anthropometric variables of the model, such as segment lengths, capsule radius, hinge rotation axes, and range of motion, which could be determined from subject body scanning or calibration [53]. The validation of such a system does require extensive testing with multi-camera optical systems, which are considered the gold standard systems for human body tracking and analysis. Although, in the proposed system, there is no need to provide these measurements to represent the movement on a 3D avatar, the length of the kinematic model segments and joints DoF can also be configured.

Another limitation of the proposed system is that it does treat the body segments independently, by means of the attached IMU sensor. In clinical and sports applications that require a very accurate kinematic assessment, the kinematic model can be used to reduce and compensate sensor tracking errors, such as noise and bias drift. The kinematics of such a model are optimally determined, for each data frame, by minimizing the difference between the measured and the estimated accelerations and angular velocities. The two most extended approaches are stochastic filtering implemented through Extended Kalman Filters (EKFs) or a constrained weighted least-square approach [51].

## 3. Experiments and Results

Different sets of tests were conducted to ensure the reliability and accuracy of the proposed system using the Custom Wearable (CW) sensors. These experiments included comparisons between our system and a digital goniometer, the Xsens DOT (XDOT) commercial sensors, and a state-of-the-art deep neural network for human body pose estimation from video, included in the Deep Stream SDK 5.0 developed by NVIDIA, which predicts and tracks in 3D up to 34 key points from images and videos. Significantly, the use of this kind of systems in medical applications related to human motion analysis is under research [55].

Movement acquisition with CW sensors was performed through the Unity3D app explained in Section 2.3. In the case of the XDOT sensors, the recordings were performed through the official Xsens Dot mobile app, using device's firmware 1.4.0 and the acquisition configuration mode "Sensor fusion Mode—Orientation (Quaternion)". Then, the Unity3D app was used for movement visualization of the acquired data at 60 Hz.

### 3.1. Validation of I2S and Body Parts Orientation Tracking

The proposed simplified kinematic model and IMU-to-segment calibration method do not consider the relative position of the sensor with respect to the bone but is useful to provide an estimation of the relative joint angle between bones. The tests conducted in this section aim to validate the I2S calibration method by comparing a static joint angle estimation with a standard digital goniometer and a dynamic joint angle estimation with a state-of-the-art deep learning approach for 3D human movement tracking, included in the Deep Stream SDK from NVIDIA.

The shortest angle between two body segments in the kinematic model presented is computed using Equation (5), where $r_a$ is the orientation of the first segment and $r_b$ is the orientation of the second segment.

$$\alpha = 2\cos^{-1}(\min(abs(r_a \cdot r_b), 1)) \cdot \frac{180}{\pi} \tag{5}$$

The shortest angle between two body segments in the computer vision approach are computed using Equation (6), where $\hat{u}$ and $\hat{v}$ are 3D vectors describing the movement of two bone segments, obtained from the tracked key points.

$$\beta = \cos^{-1}\left(\frac{\hat{u} \cdot \hat{v}}{|\hat{u}| \cdot |\hat{v}|}\right) \cdot \frac{180}{\pi} \tag{6}$$

### 3.1.1. Static Test

To verify the correctness of the proposed I2S calibration method, a custom artificial joint was built. The ends of the joint have a cylindrical shape for a better simulation of potential locations of the sensors on a real joint, but without introducing measurement errors due to clothes or skin movement. This configuration simulates a real and not ideal case placement on the human body. Since the orientation of the attached sensors may vary and they do not necessarily spin on a plane parallel to their encapsulation, it is critical to consider the initial rotation of the sensors, according to Equation (2), before estimating the shortest joint angle using Equation (5).

A digital goniometer (*Tacklife tools*) with a resolution of one tenth of a degree and a deviation of up to two tenths was used as ground truth during the experiments on static angle computation. It was attached to the right side of the artificial joint, as shown in Figure 7a.

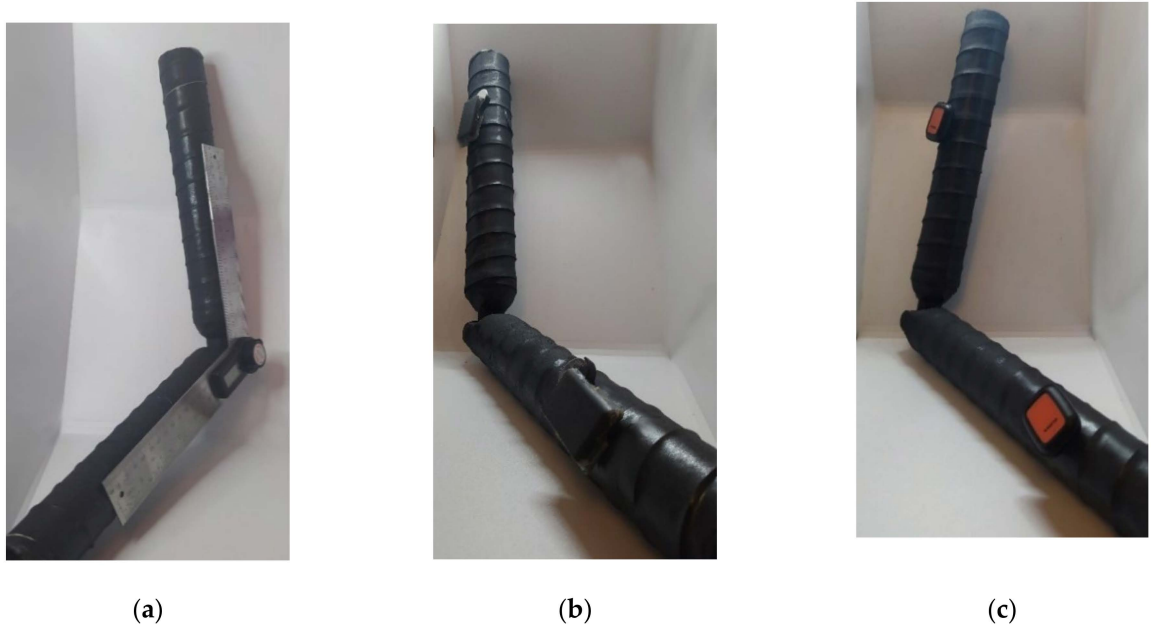

(a)                                  (b)                                  (c)

**Figure 7.** Artificial elbow joint using two cylindrical bars and a mechanic ball joint: (**a**) View of the digital goniometer from the right-side; (**b**) Custom Wearable (CW) sensors attached; (**c**) Xsens DOT (XDOT) sensors attached.

The experiment was conducted twice: using CW Figure 7b and XDOT Figure 7c sensors attached to the left side of the artificial joint after a heading reset calibration. The distance between the sensors was 45 cm in the 180° position. This position was considered as the initial pose during I2S calibration. For the tests, the digital goniometer was used to set the joint angle in different angle configurations in steps of 10°.

Table 2 shows "Mean angle" for average measured angles after 5 trial measurements of each simulated joint position and "Mean error" with standard deviation (in brackets) for the test measurements of each system. In this experiment, due to the range of movement limitation of the custom-built artificial joint, only angles up to 100 degrees were measured. The maximum detected error was within 0.60 degrees for both CW and XDOT sensors. The estimated error in this configuration was lower than 1°.

**Table 2.** Accuracy of the proposed Custom Wearable (CW) sensors set on an artificial elbow joint. Comparison with golden measurements taken from a digital goniometer set in steps of $10°$.

| Goniometer Angle (°) | CW Sensors | | XDOT Sensors | |
| --- | --- | --- | --- | --- |
| | **Mean Angle (°)** | **Mean Error (±SD) (°)** | **Mean Angle (°)** | **Mean Error (±SD) (°)** |
| 10.0 | 10.27 | $-0.27 \pm (0.08)$ | 10.37 | $-0.37 \pm (0.19)$ |
| 20.0 | 20.31 | $-0.31 \pm (0.06)$ | 20.39 | $-0.39 \pm (0.21)$ |
| 30.0 | 30.38 | $-0.38 \pm (0.07)$ | 30.46 | $-0.46 \pm (0.06)$ |
| 40.0 | 40.36 | $-0.36 \pm (0.08)$ | 40.32 | $-0.32 \pm (0.06)$ |
| 50.0 | 50.44 | $-0.44 \pm (0.07)$ | 50.17 | $-0.17 \pm (0.20)$ |
| 60.0 | 60.37 | $-0.37 \pm (0.10)$ | 60.19 | $-0.19 \pm (0.20)$ |
| 70.0 | 70.31 | $-0.31 \pm (0.12)$ | 70.17 | $-0.17 \pm (0.26)$ |
| 80.0 | 80.32 | $-0.32 \pm (0.07)$ | 80.24 | $-0.24 \pm (0.24)$ |
| 90.0 | 90.35 | $-0.35 \pm (0.14)$ | 90.15 | $-0.15 \pm (0.30)$ |
| 100.0 | 100.26 | $-0.26 \pm (0.07)$ | 100.10 | $-0.10 \pm (0.30)$ |

### 3.1.2. Dynamic Test

The aim of this experiment was to validate the feasibility of the proposed system for dynamic 3D bone orientation tracking and movement visualization. The sensors were placed in the upper-body to record elbow flexion-extension movements. Five CW sensors were employed: two sensors for each arm (*upper-arm* and *forearm*) and an additional sensor on the subject's back, for *spine* reference. The movement was recorded in video, so that it can serve to visually assess the quality of the 3D movement visualization. In addition, and with the aim to provide a more quantitative assessment, a rough estimation of the elbow joint angle obtained using Equation (3) is compared with the results obtained with the state-of-the-art deep learning optical tracking system included in the Deep Stream SDK from NVIDIA.

Figure 8a–d show four captures of the movements of the subject. The sensors were attached to the body using the official Xsens DOT straps. During the experiment, the subject was asked to bend the right arm twice, then the left arm twice, and finally both arms twice. The subject received an additional instruction for doing a quick stop when the elbow was approximately at 90 degrees. The Unity3D app was used for real-time arm movement reconstruction on a 3D avatar. Figure 8e–h show captures of this reconstruction. Videos of the subject performing the movement and its 3D reconstruction have been included as Supplementary Materials.

Figure 9a,b show graphs of the computed angle for the sequence of movements described. The blue and red plots show angle variation through time with the proposed system (Custom Wearables) and using Deep Stream (Camera Based) solution. In the first case, the angle is calculated using Equation (3) and quaternion data. In the second case, the angle is calculated using Equation (6), where $\hat{u}$ and $\hat{v}$ are 3D vectors describing the movement of the arm and the forearm, obtained from the tracked key points: shoulder, elbow, and wrist.

Red and blue plots clearly show the same trend and very close values, although the proposed CW system can track the movement more smoothly. The noisy values on the red signal are due to small variations in the inference of the z value of key points in the camera space. These small errors are inherent to the neural network, since it has been trained to track 3D points from 2D images. In addition, some variations in the amplitude of both signals can be observed, what is caused due to the fact that the neural network is estimating the root of the rotation of joints in a slightly displaced position, as it can be seen in the attached video in the Supplementary Materials. In any case, and without considering the noisy parts of the signal, the difference between both curves is never larger than 5°. Moreover, for the period in which we find elbow movement in the curves, the mean absolute error (MAE) has been measured. This value quantifies the difference in the angle estimated using our system and the camera-based approach. Right Elbow MAE for joint angle estimation in Deep Stream vs. CW analysis was 4.96° and left elbow MAE was 4.29°.

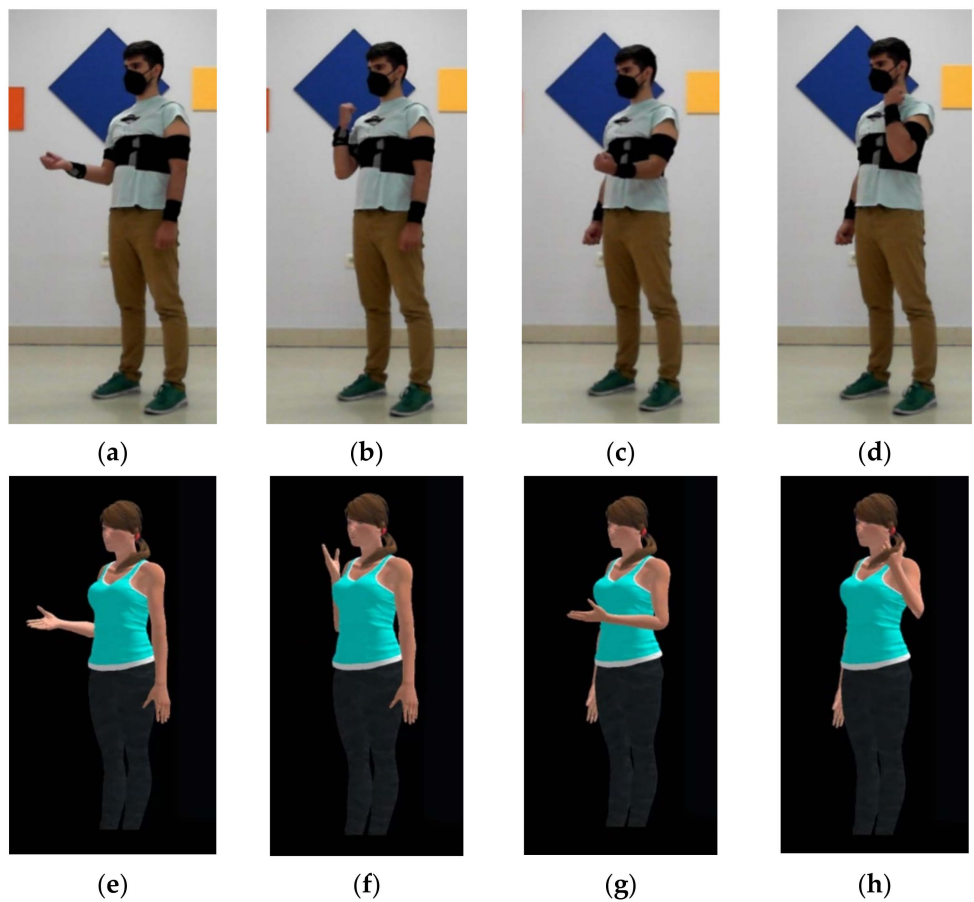

**Figure 8.** Real elbow joint instantaneous poses: (**a**) Right Elbow 90 degrees; (**b**) Right Elbow 145 degrees (**c**) Left Elbow 90 degrees; (**d**) Left Elbow 145 degrees. Corresponding poses reconstruction of real elbow joint experiment on a 3D avatar: (**e**) Right Elbow 90 degrees; (**f**) Right Elbow 145 degrees (**g**) Left Elbow 90 degrees; (**h**) Left Elbow 145 degrees.

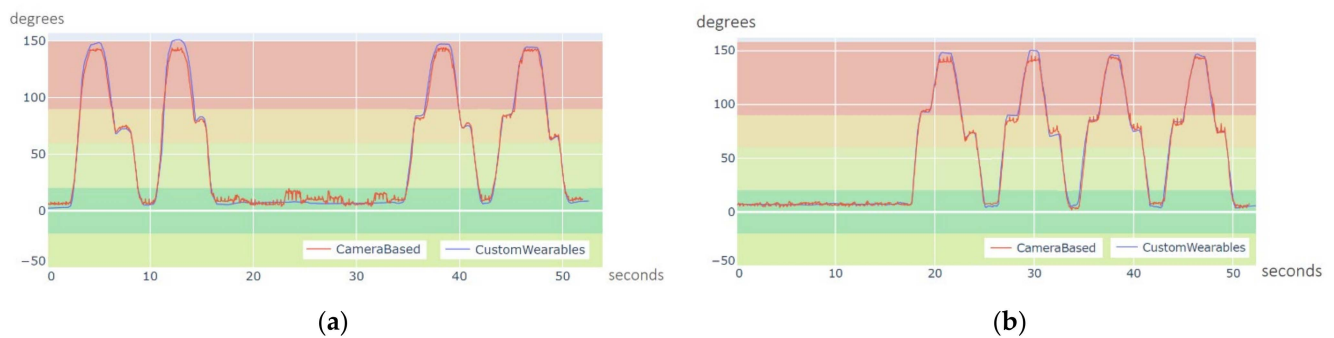

**Figure 9.** Estimated elbow joint angle for two flexion extension movements as a function of time: (**a**) Deep Stream right elbow (red) vs. CW right elbow (blue); (**b**) Deep Stream left elbow (red) vs. CW left elbow (blue).

The correlation between the angles estimated with both systems can be shown in Figure 10a,b. There is a very strong positive correlation, which implies Custom Wearables obtained a tracking of the movement comparable to that detected by the camera-based system. The value of the correlation coefficient in both cases is greater than 0.99.

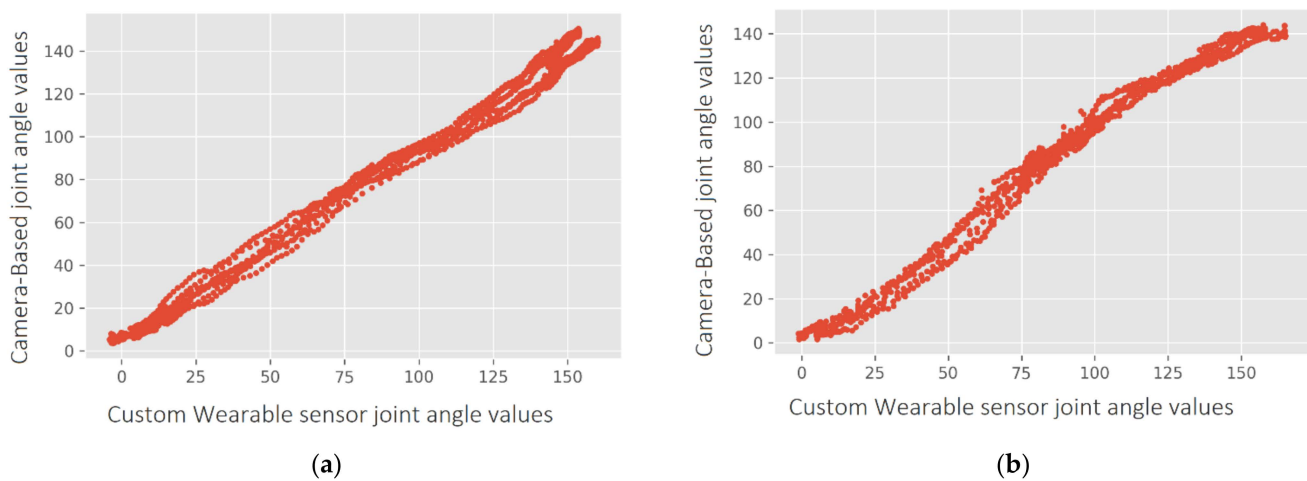

**Figure 10.** Elbow joint computed angles correlation (camera-based vs. Custom Wearables system): (**a**) Right elbow test correlation graph; (**b**) Left elbow test correlation graph.

### 3.2. Experiments Capturing Human Motion with Multiple CW Sensors

The aim of this experiment was testing our system for simultaneous movement acquisition of different body parts and fast-moving exercises using 10 and 12 CW sensors. We asked a subject to perform half jacks. The CW sensors were attached to the subject's body using the official Xsens DOT straps, as in Figure 4a.

As in experiment in Section 3.1.2, angle measurements were dynamic. The difference is that, in the previous experiments, the subject movements were done at a slower pace and only five CW and five XDOT sensors were employed. A comparison with the XDOT sensors was not performed, as the Xsens DOT system is limited to five simultaneous sensors.

Figure 11a–d show four captures of the movements of the subject, while performing the exercise. The first two captures correspond to the case with 10 CW sensors and the next two to the case with 12 CW sensors. The subject performed half jacks at high speed: as many as possible for 10 s. Figure 11e–h show captures of real-time movement visualization using the Unity3D app. Videos of the subject performing the movement and the 3D visualization have been included as Supplementary Materials.

In the first movement acquisition trial, the subject wore 10 CW sensors in total. As in Section 3.1.2 for the upper-body acquisition, the subject wore two sensors on each arm (*arm* and *forearm*) and an additional sensor on the back, for *spine* reference. Simultaneously, for the lower-body acquisition the subject wore two sensors on each leg (*thigh* and *lower leg*) and an additional sensor on *hips*. For the second movement acquisition trial, the subject wore a total of 12 CW sensors adding two CW sensors on the *feet*.

In the 10-sensor half jacks exercise, the system achieves an average rate of 50 Hz for both shoulders. In the 12-sensor half jacks exercise, the system achieves an average rate of 30 Hz for the right shoulder and 35 Hz for the left shoulder.

A comparison of Figure 11f,h illustrate the difference between the first and the second trial: in the first, feet orientation was not registered. Neither head nor hands movements have been recorded in any of the experiments.

Figure 12a shows a plot of the estimated angle for left shoulder abduction/adduction while using 10 CW sensors and Figure 12b shows an equivalent plot for the 12 CW sensors case. The first curve is smoother, which means a better acquisition resolution. The reason is that when the number of CW sensors is high, data is queued on the receiver until it can be processed. From the Unity3D interface, it is also possible to apply a spherical linear interpolation (Slerp) to quaternion data, so that the movement of the avatar becomes smoother. Figure 12c shows the resulting plot of the movement after Slerp application. This feature is only for visualization purposes, and it is not intended to estimate movement trajectories. That could be implemented using a Kalman filter, what is out of the scope of this research.

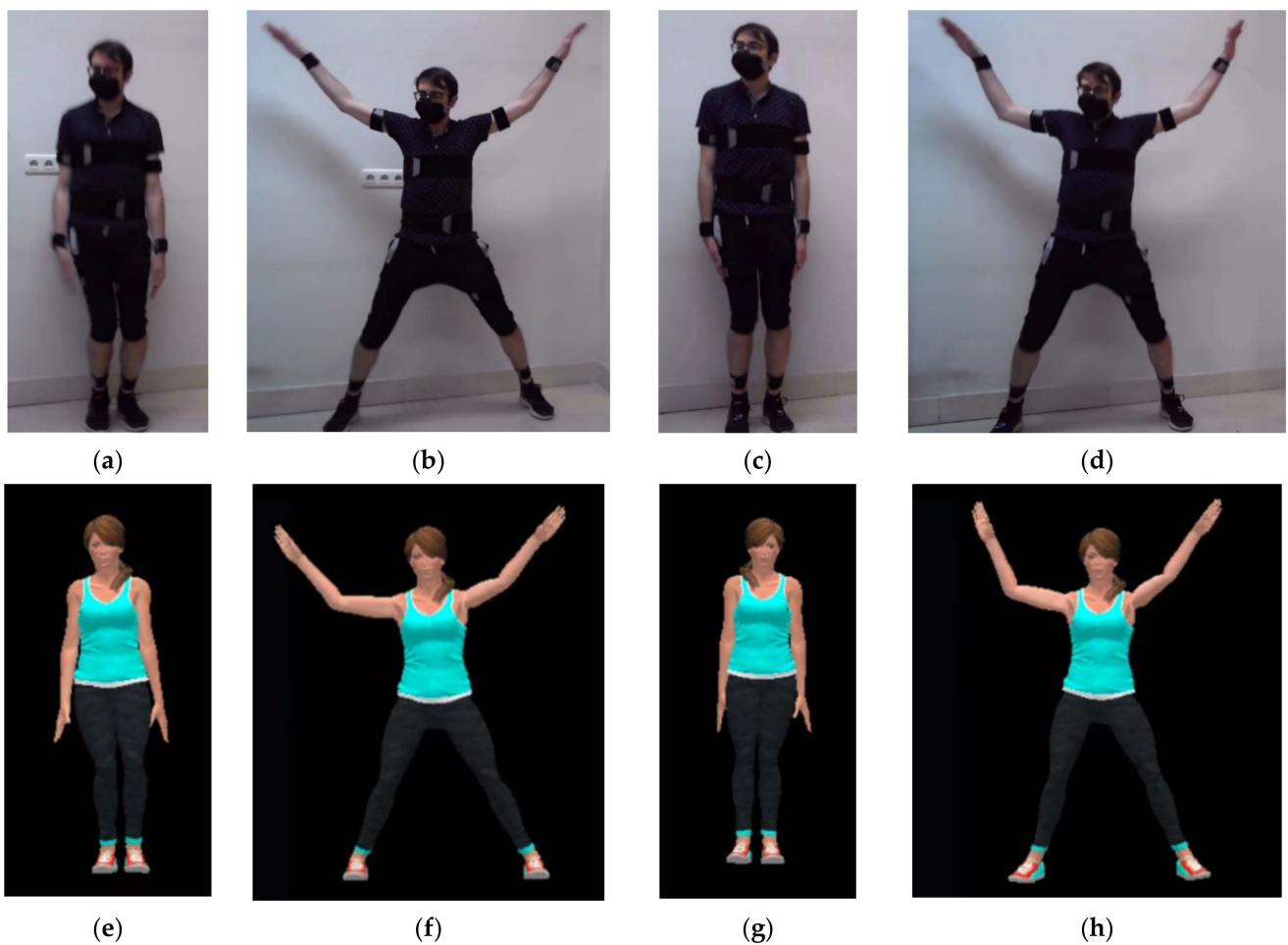

**Figure 11.** Half jacks movement acquisition with 10 and 12 CW sensors and movement reconstruction on a 3D avatar: (**a**) Initial stance with 10 CW; (**b**) Maximum Range Of Movement (ROM) position with 10 CW (**c**) Initial stance with 12 CW; (**d**) Maximum ROM with 12 CW; (**e**) Initial stance with 10 CW; (**f**) Maximum ROM position with 10 CW; (**g**) Initial stance with 10 CW; (**h**) Maximum ROM position with 12 CW.

### 3.3. Robust Communication in Bluetooth and Wi-Fi Crowded Environments

The objective behind this experiment was to test the performance of the CW and the XDOT sensors in a Bluetooth, Wi-Fi crowded environment. For the test, we used only five sensors of each kind, as this is the maximum number of XDOT sensors that can be used simultaneously. We moved across the university campus using a Bluetooth (BLE scanner) and a Wi-Fi (WiFi Analyzer) scanner apps. Figure 13a,b show a large group of Bluetooth and Wi-Fi signals detected just after we started scanning the environment in the initial location. Then, we moved towards a more crowded Bluetooth and Wi-Fi location using the mobile apps feedback: the strength of the detected signals grew or decreased depending on our position. Captures in Figure 14a,b were taken at the final chosen location. This environment includes: a group of eight Bluetooth transceivers operating in the immediate and near space and about 12 Wi-Fi access points, most of them in channels 1, 6, and 11. A more saturated color in the graph of Figure 14b, indicates a greater number of networks and congestion in that channel band.

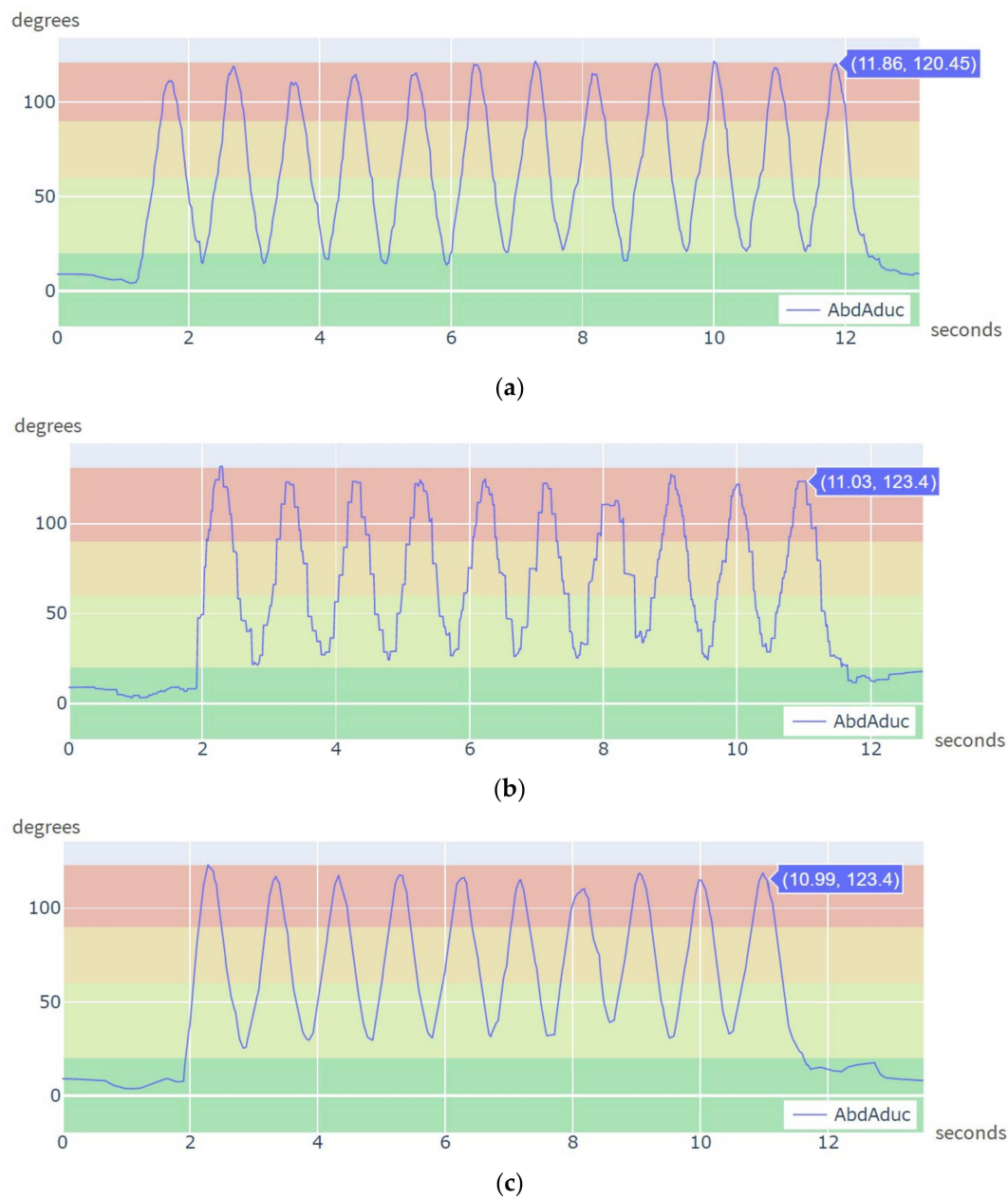

**Figure 12.** Left shoulder abduction/adduction angle during half jacks exercise: (**a**) Computed from raw-data using 10 CW sensors; (**b**) Computed from raw data using 12 CW sensors; (**c**) Same as (**b**) after spherical linear interpolation of acquired data.

For this experiment, the subject performed a frontal rising of both arms alternatively, that is, a shoulder flexion and extension. Then, flexion and extension of both elbows simultaneously. These movements were acquired using five CW sensors and then five XDOT sensors. The videos of the reconstructed 3D movements are included as Supplementary Materials.

Figure 15 shows zoomed views of the computed shoulder angle for both arms, during two flexion-extension movements. Figure 15a,b show plots of the angles of the left and right shoulders computed from the acquisition using the CW sensors, while Figure 15c,d show equivalent plots computed from the data acquired using the XDOT sensors. The *x*-axis represents the length of the movements in seconds. The movements were done twice in approximately 10 s, which cannot be considered very fast movements.

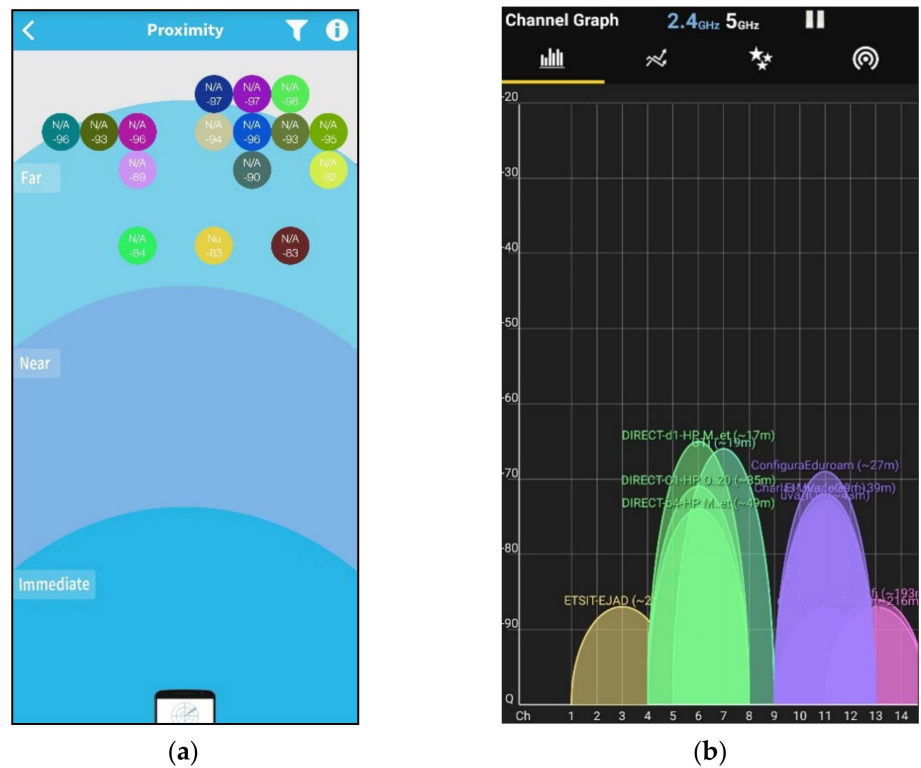

(**a**)　　(**b**)

**Figure 13.** Proximity signals for starting location. (**a**) Bluetooth signals are concentrated far away (−80 to −90 dB); (**b**) Wi-Fi signals are concentrated far away (−70 to −90 dB).

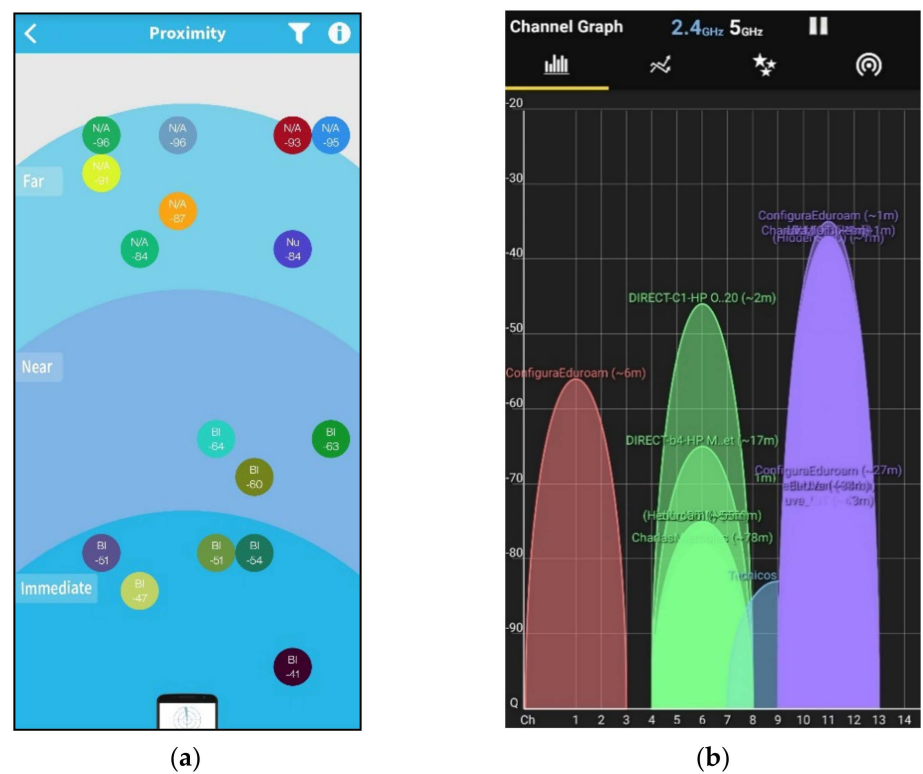

(**a**)　　(**b**)

**Figure 14.** Final chosen location signals. (**a**) Bluetooth signals in the final crowded environment (−50 to −40 dB); (**b**) Wi-Fi signals in the final crowded environment (−50 to −35 dB).

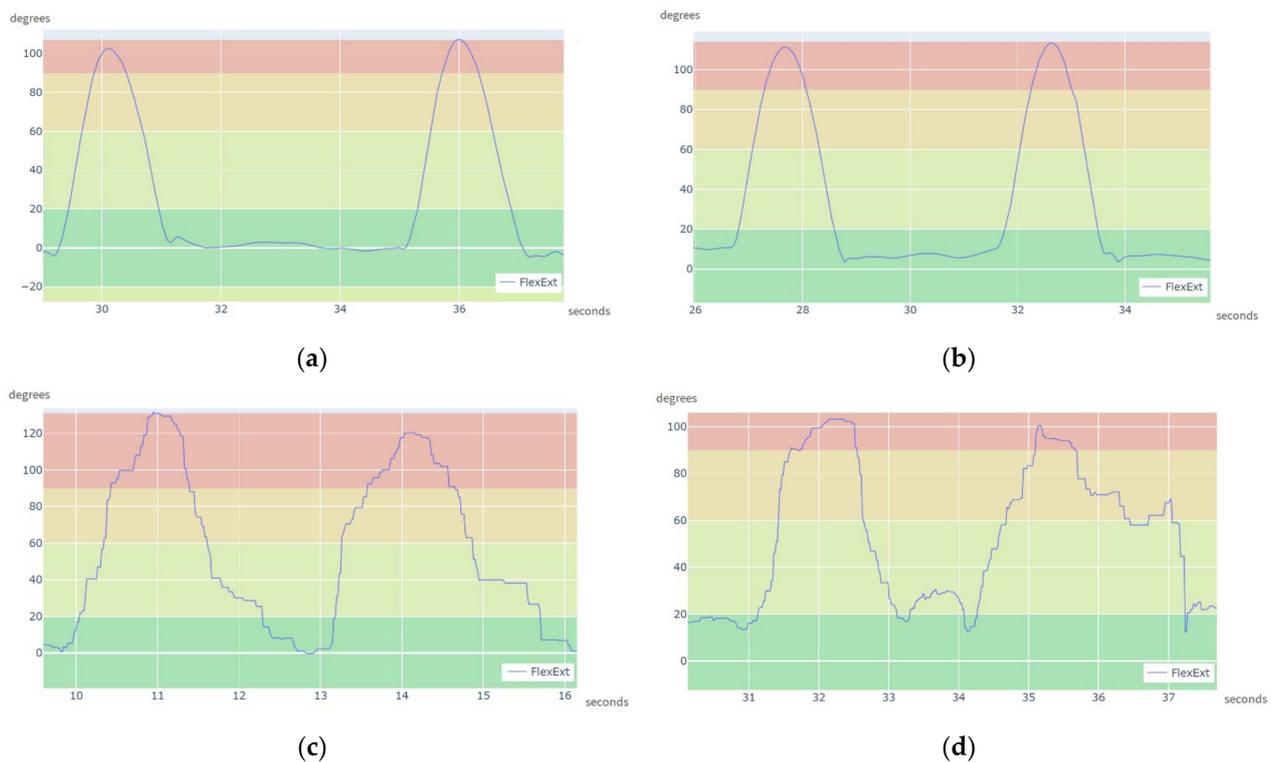

**Figure 15.** Computation angle in detail for crowded environment: (**a**) Left Shoulder angle from CW sensors data; (**b**) Right Shoulder angle from CW sensors data; (**c**) Left Shoulder angle from XDOT data; (**d**) Right Shoulder angle from XDOT data.

In Figure 15a,b the plot is smooth, which highlights that the acquisition using the CW sensors was not significantly affected by the Bluetooth and Wi-Fi interferences. On the other hand, the plots in Figure 15c,d are jagged, as a result of the data from XDOT sensors not being updated in the receiver. The XDOT sensors were significantly affected by the Bluetooth and Wi-Fi signal interferences. This behavior is observable in the graphs in Figure 15c,d.

In addition, some of the registered movements with the XDOT sensors presented synchronization problems among different sensors in this environment. Next, we better exemplify the kind of acquisition errors caused by synchronization problems by studying in detail the movement for two different acquisitions trials.

In the first trial, the subject was asked to perform alternative frontal elevation of both arms without bending the elbows. During this test, the CW accomplished an average frame rate of 48 Hz in left shoulder and 46 Hz in left elbow while the Xsens DOT achieved a rate of 17 Hz in left shoulder and 21 Hz in left elbow. Figure 16a,d show left shoulder and left elbow flexion plots and 3D movement reconstruction using the CW sensors. Figure 16b,c show left shoulder and left elbow flexion graphs using the XDOT sensors. These graphs include labels highlighting computed joint angles at precise instants in the maximum interference period (17 to 24 s) of the acquisition. Figure 16e show the movement reconstruction at second 22 and Figure 16f show movement reconstruction at second 24. Although the arm and forearm should have risen simultaneously, this is not the case. In this period, while the CW rate stays above 40 Hz, the XsensDOT rate decays drastically (below 3 Hz). These synchronization errors among nodes occurred due to the interferences and caused that the registered movement was different from the actual movement performed by the subject.

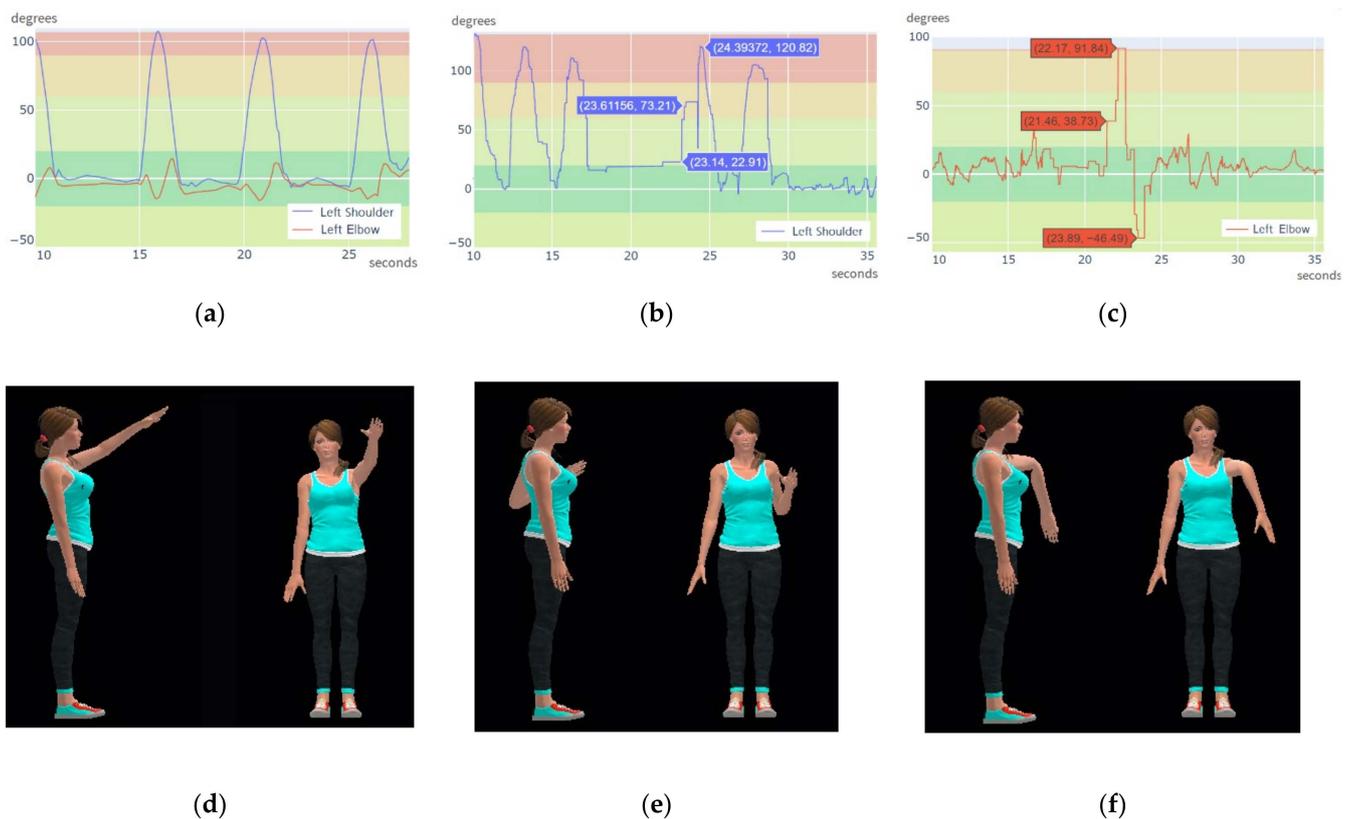

**Figure 16.** Shoulder and elbow joint angles in a 2.4 GHz crowded environment for frontal elevation of both arms: (**a**) Left shoulder and elbow flexion using CW data; (**b**) Left shoulder flexion using XDOT data; (**c**) Left elbow flexion using XDOT data. Movement reconstruction in 3D: (**d**) CW sensors; (**e**) XDOT sensors at second 22; (**f**) XDOT sensors at second 24.

In the second trial, the subject was asked to perform a simultaneous flexion and extension of both elbows. The CW accomplished an average rate of 55 Hz in both elbows while the XsensDOT achieves an average rate of 28 Hz in both elbows. Figure 17a shows the calculated joint angles of the right and the left elbow using the CW sensors. Joint tracking is smooth, and the curves are coincident, showing a synchronized movement between both arms. Figure 17d shows the reconstructed movement at second 7. Figure 17b,c show the computed angles of the left and the right elbow using the XDOT sensors data. Joint tracking is not smooth and there are synchronization errors. This is shown in Figure 17e, where the right elbow is not flexed, while the left is flexed, at second 28.5. Roughly, the right elbow movement is captured with a delay of 1 s. Figure 17f shows movement reconstruction at second 29.5 where data acquisition was synchronized again. As shown in the figures, in the worst interference period (28.5–30.5 s), the data rate in each elbow can be inconsistent, reaching only 22 Hz in left XsensDOT and down to 10 Hz in right XsensDOT. Obtaining undesired effects such as the elevation of one forearm before the other. Representative videos are included in the Supplementary Materials.

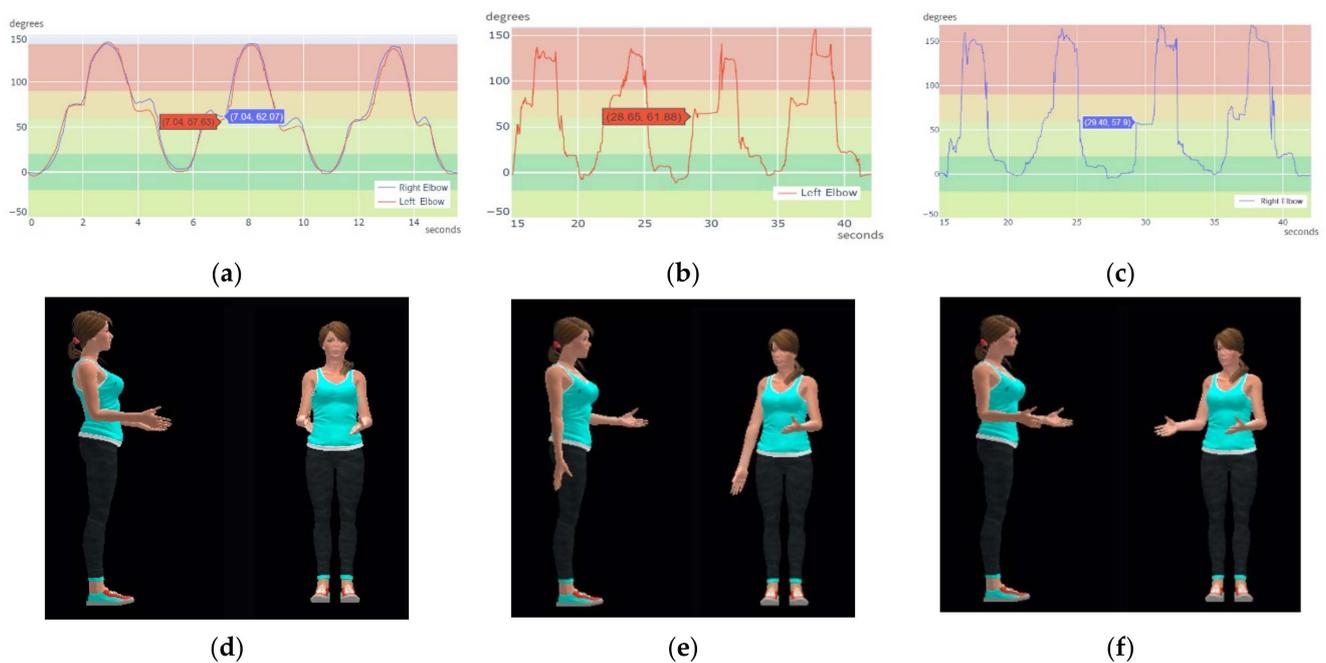

**Figure 17.** Detail of computed angle plot in a crowded environment for elbow exercise: (**a**) Left and right elbow flexion using CW sensors data; (**b**) Left elbow flexion using XDOT data; (**c**) Right elbow flexion using XDOT data. Movement reconstruction in 3D: (**d**) CW sensors; (**e**) XDOT sensors at 28.5 s; (**f**) XDOT sensors at 29.5 s.

## 4. Discussion

The set of experiments in Section 3.1 aimed to validate the proposed kinematic model and IMU-to-Segment (I2S) approach. In the test conducted in Section 3.1.1 with the proposed method and the CW sensors, an accuracy below 1° for static angle measurement was achieved, when compared with a digital goniometer. It is worth noticing that this test was performed in ideal conditions, using an artificial elbow joint, thus avoiding possible measurement errors due to clothes or skin movement. This result is comparable with previous studies using commercial IMU systems [3,8,10,16,56,57]. In the literature, different reference methods have been used. A previous study compares IMU data with a pan and tilt unit for measuring accuracy in flex-extension movements [10] and obtained errors of less than 1.05°. In another study, accuracy was measured using a custom-developed robotic device with step motors and mean error achieved for instant peak angle measurements ranged from $-0.95°$ ($\pm 0.34$) to $0.11°$($\pm 0.56$) [39]. Other study focused in wrist movement analysis, measured sensor accuracy using a robot arm and obtained an RMS error in the range from 1.1° to 1.8° [58]. The test conducted in Section 3.1.2 aimed to validate the utility of the proposed system for tracking the orientation of body parts during a dynamic movement. The tracked movement in an elbow a flex-extension exercise was compared to Deep Stream, a state-of-the-art deep learning optical tracking system. Results showed a difference in the computed joint angle always smaller than 5° between both systems, obtaining a MAE for Right Elbow of 4.96° and for Left Elbow of 4.29°. The value of the correlation coefficient in both cases is greater than 0.99.

Other research works use as reference optoelectronic or camera-based systems [8,33,39]. Previous works study the reliability and accuracy of custom developed sensors using robotic arms [12,39] and optoelectronic systems [11,14,17].

The set of experiments in Section 3.2 were devised to analyze the performance of the proposed system for multiple body parts orientation tracking, using 10 and 12 CW sensors simultaneously. The movements in these tests were much faster than in previous tests. The results show that the CW sensors can smoothly track the half jacks exercise using 10 CW sensors at 50 Hz. Using 12 CW sensors, and thus tracking two more body parts (e.g., feet), is still feasible, but at 35 Hz. Other studies in the literature using custom sensors employed only one to three sensors simultaneously [14,17,18,39]. The purpose of these tests is not accurate shoulder joint angle measurement. A more complex biomechanical model of the shoulder, considering scapular protraction and retraction movements could be developed on top of the proposed system [59,60].

The first objective of this work was to prove the feasibility of building an affordable state-of-the-art custom do-it-yourself solution for multiple body parts orientation tracking. The achievement of this objective was demonstrated in those experiments. The second objective of this work was to prove that do-it-yourself solutions can also improve some of the features of entry-level commercial systems to some extent. Table 3 subsumes the main features of the proposed system and compares it with entry-level and high-end commercial solutions.

**Table 3.** Features of the proposed custom do-it-yourself system and other entry-level and high-end commercial solutions.

| | Do-It-Yourself | Entry-Level | | High-End | |
|---|---|---|---|---|---|
| **Wearable Solution** | **CW** | **Xsens DOT** | **Notch** | **Xsens MTwAwinda** | **Perception Neuron PRO** |
| Real-time max. simultaneous IMUs | 10 | 5 | 6 | 20 | 17 |
| Sampling rate (output) | 60–50 Hz | 60 Hz | 40 Hz | 120 Hz–60 Hz (5–20 nodes) | Up to 96 Hz |
| Battery Life | ≥3 h 90 mAh | ≥5.5 h 70 mAh | ≥6 h | ≥6 h | ≥3.5 h |
| Hardware cost (without taxes) | 250 € (10 units + dongle) | 495 € (5 units) | 429 $ (6 units) | 4390 € (6 units) | 3500 $ (17 + 1 units) |
| Communication range (not open space) | <10 m | <10 m | <10 m | 20–10 m | >20 m |

The number of simultaneous sensors that can be used for body tracking in high-end commercial solutions such as Xsens MTwAwinda [31] and PerceptionNeuron is around four times larger than in entry-level solutions such as XsensDOT and Notch, which are capable of connecting a maximum of five or six simultaneous nodes. The latter do not reach the number of simultaneous sensors of professional high-end solutions, due to their use of Bluetooth (either BLE or BT 5.0) for compatibility with mobile devices. In the custom proposed system, the number of sensors is in between and doubles the number of sensors that can be used with a commercial entry-level system. In relation to the sampling rate, the proposed solution has an equivalent performance than entry-level solutions, which is enough for the use cases considered in this work. High-end commercial solutions significantly increase the sampling rate. Significantly, in sports applications a sampling rate higher than 60 Hz is recommended, due to the speed of the subject movements [36]. In custom-developed system, on-board storage could be added to increase the sampling rate for these specific applications. Regarding battery life, all solutions in the table have a reasonable autonomy for many medical use cases, where the sessions do not usually exceed 1 h.

The average cost of the system in the proposed solution is significantly lower than in the case of high-end commercial solutions and the cost of each sensor unit is reduced by a factor of about four compared to entry-level commercial solutions. An important part of the hardware cost is related to the internal IMU. In the proposed system the chosen IMU is the BNO080, same as in [12]. Other commonly used IMUs in previous works are MPU6050 [61], MPU9150 [39], MPU9250 [62] or BNO055 [17,18]. Remarkably, the costs in the table do not consider software costs. In the proposed system, the firmware of the sensor nodes was developed using the FreeRTOS and higher software layers using a no-cost personal version of Unity3D game engine. The developed tools can represent the movement in a 3D avatar in real time. In entry-level systems, only a free API for raw data acquisition is provided. These systems do not include any software for human movement visualization during data acquisition. In high-end solutions, such as Xsens MTwAwinda extra licenses must be paid in a software as a service model. The software can be used not only to represent the movement on an avatar, but also to compute a large set of parameters related to human movement analysis (e.g., gait analysis and ergonomic assessment).

The maximum communication range of the proposed solution is within 10 m. In entry-level systems, where the sensors are intended to be used with a mobile device, the maximum communication range is equivalent. On the high-end solutions, the use of a powerful transceiver device with an antenna can double the communication range.

The set of experiments in Section 3.3 aimed to validate the use of the proposed custom communication protocol and channel hopping strategy for improving communication features over entry-level commercial systems. In these tests we compared the performance of the CW and the XDOT sensors in a crowded Bluetooth and Wi-Fi 2.4 GHz environment. These environments are becoming increasingly frequent, due to the advent of internet of things. Our tests demonstrated that entry-level commercial sensors using Bluetooth suffer from acquisition errors due to network interferences in these environments. This means that the registered movements could be different from the real movements the subject performed. This channel hopping strategy is neither included in the entry-level nor in the most affordable solutions, and in our field of application, the inclusion of this strategy in an accessible solution allows us to expand its use to any environment and user.

The results show that our wireless communication approach is more robust to possible interferences in the 2.4 GHz band. These tests highlight that if a system does not include a mechanism to prevent this kind of interferences, the system will not work, and the acquired data might not reflect the real movement performed by the subject. A real-time 3D movement visualization tool like the one proposed in this work can help to detect common issues during motion acquisition. The two entry-level solutions reviewed have no special communication mechanism detailed in their specifications, to avoid interferences in crowded environments. High-end solutions also avoid using Bluetooth for their communication procedure, designing their proprietary protocols with interference avoiding mechanisms on the ISM 2.4 GHz band.

In this research paper, an affordable custom-designed IMU-based wireless system, aimed at multiple body parts orientation tracking, was presented. The proposed solution addresses some of the main limitations of previous custom IMU-based systems in the literature and entry-level commercial solutions. The system can use 10 sensors at 50 Hz simultaneously and achieves a robust wireless communication in 2.4 GHz Wi-Fi crowded environments, where the use of entry-level commercial solutions is unfeasible. This work also provides a bottom-up description on the hardware, tools and mathematical operations employed, including a Unity3D app for 3D movement visualization.

The developed system presents some limitations. The maximum number of simultaneous CW sensors and data sampling frequency can be further improved. Future research will explore the use of data compression and packet buffering techniques in this regard. The sampling rate achieved in the proposed system (1–10 sensors at 60–50 Hz) is not very high. Nonetheless, it is enough for some medical applications. As a matter of fact, previous studies find the applicability of systems with equal or even lower frequency. These studies are targeted to motion estimation, but down sample inertial sensor data from commercial or custom IMUs at a final frequency of 50 Hz [43,63], use the system for rehabilitation supervision at a sampling rate of 40 Hz [36], or perform ambulatory human joint motion monitoring at 50 Hz [64]. Other limitations of the system regarding the kinematic model have been commented in Section 2.4. Finally, no experiments have been conducted to measure the impact of the internal drift of the IMU sensor. Drift can limit the usability of the system, and this has been previously evaluated in other works, such as [12]. In the future, we aim to evaluate the use of the proposed system in more specific medical and industrial applications that require longer capture times.

## 5. Conclusions

The proposed system proves the feasibility of developing an affordable custom system that can address some of the main limitations of previous custom and entry-level commercial solutions at a fraction of the cost. The system can use up to 10 IMU-based, wirelessly connected sensors at 50 Hz in the 2.4 GHz band but using channel switching strategies to achieve reliable data acquisition in Bluetooth and Wi-Fi crowded environments. This custom do-it-yourself system can be used for real-time human movement acquisition and visualization on a 3D avatar.

The potential applications of this solution include those in which fast calibration or real-time 3D movement visualization is a requirement, but full-featured kinematic body analysis is not demanded. Examples of tools that can be developed on top of the proposed system include in-home rehabilitation tools, therapy follow-up assessment in the natural environment of the patient, or human activity recognition solutions. In addition, it can be further developed to include more complex kinematic models of movement for specific medical assessment applications.

Despite the potential usefulness of inertial measurement units, neither commercial nor custom systems are yet widely employed in real applications. We can foresee that the development of solutions as the one proposed in this work will increase. Eventually, this will popularize the use of custom IMU-based wearable proposals in human motion related applications.

**Supplementary Materials:** The following are available online at https://www.mdpi.com/article/10.3390/s21196642/s1, Section 3.1.2. Dynamic test: Video S1: ElbowFlexion_Avatar.mp4, Video S2: ElbowFlexion_DeepStream.mp4, Video S3: ElbowFlexion_RealSubject.mp4; Section 3.2. Experiments capturing human motion with multiple CW sensors: Video S4: 10_CustomSensors_JumpingJacks.mp4, Video S5: 10_CustomSensors_JumpingJacks_Camera.mp4, Video S6: 12_CustomSensors_JumpingJacks.mp4, Video S7: 12_CustomSensors_JumpingJacks_Camera.mp4; Section 3.3. Robust communication in Bluetooth and Wi-Fi crowded environments: Video S8: CWsensors_ArmsRise.mp4, Video S9: CWsensors_ElbowFlexion_interferences.mp4, Video S10: XsensDOTsensors_ArmsRise.mp4, Video S11: XsensDOTsensors_ElbowFlexion.mp4.

**Author Contributions:** Conceptualization, M.M.-Z. and J.G.-A.; methodology, M.M.-Z. and F.J.D.-P.; software, J.G.-A., D.O.-P. and M.M.-Z.; hardware, J.G.-A.; validation, J.G.-A., H.J.A., D.G.-O. and M.M.-Z.; data curation, J.G.-A.; writing—original draft preparation, J.G.-A., M.M.-Z.; writing—review and editing, H.J.A., F.J.D.-P., D.G.-O. and M.M.-Z. All authors have read and agreed to the published version of the manuscript.

**Funding:** This research has been partially funded by a research contract with IVECO Spain SL and by the Department of Employment and Industry of Castilla y León (Spain), under research project ErgoTwyn (INVESTUN/21/VA/0003).

**Institutional Review Board Statement:** The study was conducted according to the guidelines of the Declaration of Helsinki, and approved by the Institutional Review Board (or Ethics Committee) of "CEIm ÁREA DE SALUD VALLADOLID ESTE" (protocol code PI 21-2341).

**Informed Consent Statement:** Informed consent was obtained from all subjects involved in the study.

**Conflicts of Interest:** The authors declare no conflict of interest.